\documentclass{pnastwo}

\usepackage{graphicx}%[dvips]{graphicx}
\usepackage{amssymb,amsfonts,amsmath,bm,dsfont}
\usepackage[usenames]{color}

\url{www.pnas.org/cgi/doi/10.1073/pnas.0709640104}
\copyrightyear{2008}
\issuedate{Issue Date}
\volume{Volume}
\issuenumber{Issue Number}
\begin{document}
\title{Nematics, knots and non-orientable surfaces}
\author{Thomas Machon\affil{1}{Centre for Complexity Science, University of Warwick, Coventry, CV4 7AL, United Kingdom}
\and
Gareth P. Alexander\affil{1}{}\affil{2}{Department of Physics, University of Warwick, Coventry, CV4 7AL, United Kingdom}}

\contributor{Submitted to Proceedings of the National Academy of Sciences of the United States of America}
\maketitle
\begin{article}
\begin{abstract} 
Knots and knotted fields enrich physical phenomena ranging from DNA and molecular chemistry to the vortices of fluid flows and textures of ordered media. Liquid crystals provide an ideal setting for exploring such topological phenomena through control of their characteristic defects. The use of colloids in generating defects and knotted configurations in liquid crystals has been demonstrated for spherical and toroidal particles and shows promise for the development of novel photonic devices. Extending this existing work, we describe the full topological implications of colloids representing non-orientable surfaces and use it to construct torus knots and links of type ($p$,2) around multiply-twisted M\"obius strips. 
\end{abstract}


\dropcap{C}ontrolling and designing complex three-dimensional textures in ordered media is central to the development of advanced materials, photonic crystals, tunable devices or sensors and metamaterials \cite{poulin97,photo1,lapointe09,toron,araki11,lin11,ravnik11,lavrentovich11,honglawan13,musevic13}, as well as to furthering our basic understanding of mesophases \cite{chen09,lub1,bryan}. Topological concepts, in particular, have come to play an increasingly significant role in characterizing materials across a diverse range of topics from helicity in fluid flows \cite{moreau61,moffatt69} and transitions in soap films \cite{goldstein10}, to molecular chemistry \cite{herges06}, knots in DNA \cite{marenduzzo09}, defects in ordered media \cite{mermin,gareth}, quantum computation \cite{kitaev03,nayak08} and topological insulators \cite{hasan10}. Topological properties are robust, since they are protected against all continuous deformations, and yet flexible for the same reason, allowing for tunability without loss of functionality. 

Some of the most intricate and interesting textures in ordered media involve knots. Originating with Lord Kelvin's celebrated `vortex atom' theory \cite{thomson67}, the idea of encoding knotted structures in continuous fields has continued in magnetohydrodynamics \cite{woltjer58}, fluid dynamics \cite{moffatt69}, high energy physics \cite{faddeev97,sutcliffe07,witten89}, and electromagnetic fields \cite{ranada92,irvine08}, and has seen recent experimental realizations in optics \cite{dennis10}, liquid crystals \cite{knots} and fluid vortices \cite{kleckner13}. Tying knots in a continuous field involves a much greater level of complexity than in a necktie, or rope, or even a polymer or strand of DNA. In a field, the knot is surrounded by material that has to be precisely configured so as to be compatible with the knotted curve. However, this complexity brings its own benefits, for the full richness of the mathematical theory of knots is naturally expressed in terms of the properties of the knot complement -- everything that is {\em not} the knot. In this sense, knotted fields are ideally suited to directly incorporate and experimentally realize the full scope of modern knot theory. 

Liquid crystals are orientationally ordered mesophases, whose unique blend of soft elasticity, optical activity and fluid nature offer a fertile setting for the development of novel metamaterials and the study of low dimensional topology in ordered media. Much of the current focus centers on colloidal systems -- colloidal particles dispersed in a liquid crystal host -- which have a dual character. On the one hand, the liquid crystal mediates long-range elastic interactions between colloids, furnishing the mechanism for formation of colloidal structures and metamaterials \cite{poulin97,photo1,lapointe09,musevic13}. On the other hand, the colloids, through anchoring conditions imposed by their surfaces, generate defects in the liquid crystal and so serve to induce and manipulate its topological properties. For instance, multiple colloids exhibit a variety of entangled defect configurations \cite{ravnik07,copar12}, equally interesting states without defects \cite{tkalec09}, and can even be manipulated so as to form arbitrary knots and links \cite{knots,jampani11}. More recently, a significant advance has seen the fabrication of colloids with different topology \cite{torus} -- tori up to genus five -- verifying experimentally the relation between particle topology and accompanying defect charge, and advancing a program to obtain topological control of materials through topological design. While the phenomena displayed by these systems is indeed rich, as surfaces these colloids (spheres, tori, \textit{etc.}) all represent closed, orientable surfaces.

In this article we extend these ideas to provide a complete topological characterization of all compact colloidal surfaces in a liquid crystal host. {\em Non}-orientable surfaces fully exploit the non-orientable nature of liquid crystalline order \cite{ball11}. We show that the topology of non-orientable surfaces enforces the creation of topologically protected disclination lines. By varying the embeddings of the surface we exploit this topology to create metastable disclination loops in the shape of $(p,2)$ torus knots and links, for any $p$, around multiply-twisted M\"{o}bius bands. Through this combination of geometry and topology we elucidate a natural setting for the creation and control of complex knotted fields and the integration of mathematical knot theory into experimental science. 

\section{Colloids, Surfaces and Topology}

Liquid crystals are typically composed of long, thin, rodlike molecules, which align in the nematic state along a common direction ${\bf n}$ called the director. This orientation is line-like rather than vectorial, meaning ${\bf n}\sim -{\bf n}$ and the local orientation takes values in the real projective plane $\mathbb{RP}^2$, the ground state manifold for nematics \cite{mermin,gareth}. The director field varies smoothly everywhere except on points or lines of discontinuity, known as topological defects, whose nature is captured by the way the order changes in their vicinity. Line defects, called disclinations, are characterized by the behavior on small loops around the singular line and so may be classified by the fundamental group of the ground state manifold $\pi_1(\mathbb{RP}^2)=\mathbb{Z}_2$. Unique to liquid crystals, these defects reflect the non-orientability of nematic order with the molecules undergoing a $\pi$-rotation upon encircling the disclination. Similarly, point defects, referred to colloquially as hedgehogs, are classified by the second homotopy group $\pi_2(\mathbb{RP}^2)=\mathbb{Z}$. This simple classification is augmented by the topological content of interactions between defects (characterized in terms of the action of $\pi_1$ on the higher homotopy groups, or by various Whitehead products \cite{poenaru77}) so that, in particular, point defects of equal but opposite strength are equivalent (meaning freely homotopic) in the presence of a disclination and, whilst there is only one non-trivial element of $\pi_1(\mathbb{RP}^2)$, disclinations that close up into loops fall into four distinct homotopy classes \cite{janich}. Loosely, these may be thought of as corresponding to whether the linking number with other defects is even or odd and whether the loop carries an even or odd hedgehog charge \cite{gareth}. 

Defects can be induced and manipulated by immersing colloidal particles in the liquid crystal. This arises through the incompatibility of anchoring conditions on the particle surfaces with the alignment imposed by the cell boundaries, or at large distances. The topological type of this incompatibility, or obstruction, can be associated with elements of the homotopy groups $\pi_k(\mathbb{RP}^2)$ so that different surfaces can (loosely) be thought of as generating different types of defects. The nature of the obstruction depends on the anchoring conditions at the surface, but in the most common case of normal, or homeotropic, anchoring (as we consider here) it depends only on the topology of the colloid's surface. The classification theorem of surfaces \cite{Donaldson} is a classic result of two-dimensional topology which states that any compact surface can be classified up to homeomorphism by its genus, orientability and number of boundary components. The genus is equal to the number of holes or handlebodies possessed by a surface; for example, a torus has genus one and a sphere genus zero. Orientability implies a consistent choice of normal vector can be made on a surface. The one-sided M\"{o}bius strip is the classic non-orientable surface; any normal vector on the strip will be flipped by going around the strip once, forbidding a consistent choice of surface normal. Finally, the number of boundary components is simply the number of distinct connected components in the surface boundary, \textit{e.g.} a disk has one boundary component and a torus has none. 

While this is a complete topological classification of surfaces in an abstract setting, for applications they must also be embedded (no self-intersections) into ordinary three-dimensional space, $\mathbb{R}^3$. Different embeddings are interesting in their own right -- the whole of knot theory concerns embeddings of a circle into $\mathbb{R}^3$ -- but they do not affect the homotopy class of the defect necessitated in the bulk, which we focus on first. Thus, with the classification of surfaces in mind it is natural to ask what topological implications each type of surface has for accompanying defects in the surrounding liquid crystal. The complete classification, summarized in Fig.~1, naturally separates into four classes of surfaces; orientable or non-orientable and closed or with boundary. 

Closed, orientable surfaces are known to induce defects corresponding to the element $1-g=\chi/2$ of $\pi_2 (\mathbb{RP}^2)$, where $g$ is the genus of the surface and $\chi$ is the Euler characteristic~\cite{milnor,torus}. Briefly, this relation comes through computing the degree of the Gauss map of the surface. The Gauss map, $\mathcal{G}$, of a surface, $X$, is a map $\mathcal{G}: X \to S^2$ that sends every point of the surface to the direction of the surface normal at that point. For orientable surfaces with normal anchoring, the director can be given the orientation of the Gauss map, so that ${\cal G}$ describes precisely the molecular orientation at the surface. The degree of this map -- the number of times every point on $S^2$ is visited, counted with sign -- is a homotopy invariant~\cite{milnor} characterizing the type of defect that the surface generates \cite{torus}. Although experimentally the same surface can produce seemingly different defects, they are always characterized by this same element of $\pi_2 (\mathbb{RP}^2)$. For instance, spherical colloids can nucleate either a point defect \cite{poulin97} or disclination loop \cite{terentjev95} but the loop can always be shrunk continuously into a point \cite{stark01}, so that it is more properly classified by $\pi_2 (\mathbb{RP}^2)$. More generally, orientable surfaces can never generate elements of $\pi_1(\mathbb{RP}^2)$, \textit{i.e.} disclinations, as a topological requirement. Their orientability ensures that any disclination loops formed can always be removed in pairs or shrunk into points. 

Closed, non-orientable surfaces cannot be embedded in $\mathbb{R}^3$ without self-intersection, meaning that a true representation of any of these (\textit{e.g.} the real projective plane or the Klein bottle) in a liquid crystal is not possible. For this reason, we do not consider closed, non-orientable surfaces any further. 

Orientable surfaces with boundaries have trivial topological implications for the surrounding liquid crystal. Their orientability, as in the closed case, forbids them from generating elements of $\pi_1(\mathbb{RP}^2)$ as a topological necessity. In addition, since they have a boundary, they cannot generate any elements of $\pi_2(\mathbb{RP}^2)$. We argue as follows. Closed, orientable surfaces separate space into an inside and an outside\footnote{We consider only ``tame'' embeddings such that the piecewise linear Sch\"{o}nflies theorem applies.} and they generate point defects in {\it both} regions. If one cuts a hole in the surface, creating a boundary component, then these defects can be combined to leave a defect-free texture. Cutting more holes in the surface will not change things as one could always make the texture on the new hole identical to the surface normal that was removed. 

\begin{figure}[t]
\centering
\includegraphics[width=0.49\textwidth]{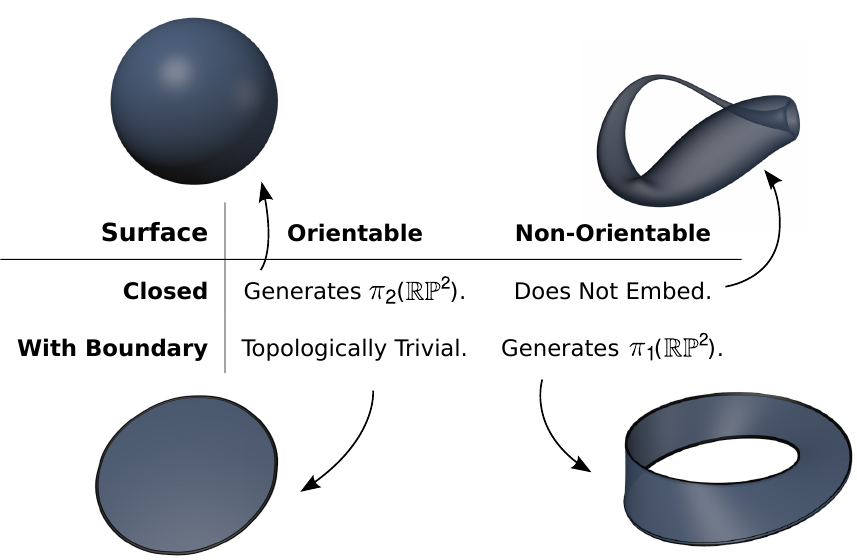}
\caption{Topological characterization of compact surfaces with homeotropic boundary conditions, embedded in a three-dimensional, nematic liquid crystal. Closed orientable surfaces generate elements of $\pi_2(\mathbb{RP}^2)$, equal to one minus the genus of the surface. Non-orientable surfaces with boundary generate the non-trivial element of $\pi_1(\mathbb{RP}^2)$, which forces the nucleation of disclination lines in the bulk.}
\label{tab:1}
\end{figure}

Non-orientable surfaces with boundary necessarily generate a non-trivial element of $\pi_1(\mathbb{RP}^2)$. A generalization of the Gauss map to non-orientable surfaces $\overline{{\cal G}}:X \to \mathbb{RP}^2$ assigns to every point of the surface the line element (point in $\mathbb{RP}^2$) corresponding to the direction of the (unoriented) surface normal\footnote{It is more common, and more useful in general, to think of the non-orientable version of the Gauss map as a map to the Grassmannian $\text{Gr}_2(\mathbb{R}^3)$ -- see, {\it e.g.} Milnor JW, Stasheff JD (1974) {\em Characteristic Classes} (Princeton Univeristy Press, Princeton), pp 55-70 -- but since $\text{Gr}_2(\mathbb{R}^3)$ is canonically homeomorphic to $\mathbb{RP}^2$ there is no loss in our discussion.}. Now consider the loop space of our surface, $\Omega X$. Composition with the non-orientable Gauss map $\overline{{\cal G}}$ creates a set of representatives of $\pi_1(\mathbb{RP}^2)$. If $X$ is non-orientable then there must be at least one map in this set which represents the non-trivial element (and generates a disclination). Suppose there was no such map, then every map in $\overline{{\cal G}}[\Omega X]$ could be lifted from $\mathbb{RP}^2$ to $S^2$ and we would have created an orientable Gauss map, $\mathcal{G}$, implying the surface is orientable, a contradiction. The disclinations created in this way \textit{must} entangle the surface, since any disk spanning a non-orientable loop on the surface must be pierced by the defect. 

With closed, orientable surfaces, different elements of $\pi_2(\mathbb{RP}^2)$ could be generated according to the genus of the surface. In the same way, we might ask if different non-orientable surfaces with boundary can act to generate different kinds of disclination loops in the bulk liquid crystal. Although there is only one non-trivial element of $\pi_1(\mathbb{RP}^2)$, there are four distinct homotopy classes of disclination loops \cite{janich,gareth} (the four types may be thought of as corresponding to even/odd linking number with other disclinations and even/odd hedgehog charge). We must, therefore, determine which of these is generated by the surface. We cannot force the existence of linked loops in a path connected domain. If we have a connected domain then a single disclination loop can go through all the non-trivial cycles created by the surface. Since we will then have only one disclination, it cannot be linked. Furthermore, it must have zero hedgehog charge. The surface simply acts to smoothly align the director field in some region of space so that the texture is equivalent to one containing a lone disclination loop, without the colloid but with the same director orientation in its place, which must have zero charge. Thus all non-orientable surfaces with boundary, independent of their topological type, have the same topological implication for the liquid crystal, in contrast to the closed, orientable case. These four classifications are summarized in Fig.~1. 

\section{M\"{o}bius Strips and Knotted Defects}

Non-orientability of the surface enforces the existence of a disclination loop but it leaves open the precise form of the defects and their equilibrium configuration. These are determined by energetics and by the nature of the embedding of the surface. While surfaces with boundary are unavoidably two-dimensional, experimentally realizable, but topologically equivalent, surfaces may be constructed by using thin material with homeotropic boundary conditions on the faces and planar anchoring on the thin edges, as shown schematically in Fig.~2. In this way a colloid with varying surface anchoring conditions can be made to faithfully represent a two-dimensional surface. Such boundary conditions are essential to mimic non-orientable surfaces and ensure the topological properties we describe -- fully homeotropic boundary conditions simply replicate an orientable torus. Modern fabrication techniques allow for the manufacture of such exotic surfaces \cite{torus}. 

As the M\"{o}bius strip is the prototypical non-orientable surface -- \textit{all} non-orientable surfaces contain the M\"{o}bius strip as a subset -- it serves as an elementary guide to the behavior of non-orientable surfaces in liquid crystals. A M\"{o}bius strip with homeotropic boundary conditions will generate a non-trivial element of $\pi_1(\mathbb{RP}^2)$ as one passes around the strip and hence must be threaded by a disclination loop of zero hedgehog charge, entangling the surface. The shape of disclination loops around colloidal particles is governed largely by the requirement to minimize the distortion energy of the surrounding director field. A simple heuristic for this can be constructed as follows. As one passes through a disclination line a rotation of approximately $\pi/2$ is induced in the director. This rotation mediates the transition from the surface normal orientation to that of the far-field and will best minimize the distortion in the director field when it is concentrated along those parts of the surface where the local anchoring and far-field directions are perpendicular. Since we consider colloids with homeotropic anchoring this simply gives the requirement that
\begin{equation}
\textbf{S}_n \cdot \textbf{n}_0 = 0 ,
\label{eq:1}
\end{equation}
along the disclination, where $\textbf{S}_n$ is the surface normal and $\textbf{n}_0$ is the far-field director. For a spherical particle in a uniform far-field, this predicts that the disclination will lie on a great circle in a plane perpendicular to this far-field direction, which is the observed position of Saturn ring defects \cite{terentjev95}. Likewise the twisted shape of disclinations around spherical colloids in a cholesteric \cite{lintuvuori10,jampani11} is correctly predicted by the same heuristic. 

\begin{figure}[t]
\centering
\includegraphics[width=0.49\textwidth]{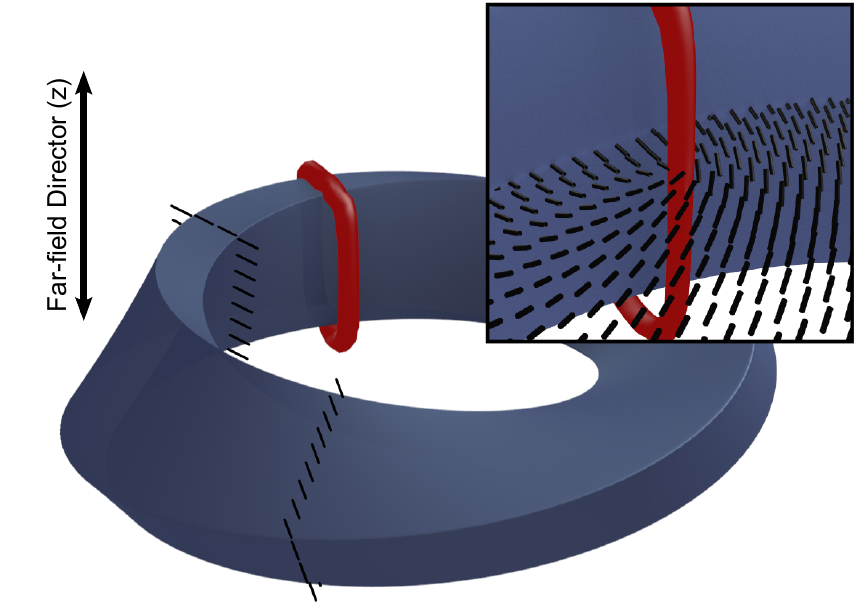}
\caption{Simulation results of a M\"{o}bius strip with homeotropic boundary conditions on the flat faces, shown schematically in black along the strip. Note the planar anchoring on the short edge required to correctly represent a two-dimensional surface. The two-dimensional surface can then be thought of as living in the center of the strip. The defect line (red) is clearly visible entangling the strip, as predicted by \eqref{eq:1}. The inset shows the local director profile near the defect, showing the +1/2 twist disclination on the inside of the strip; there is a corresponding -1/2 twist profile on the outside.}  
\label{fig:mob2}
\end{figure}

The preferred defect configuration for a M\"{o}bius strip can be found by numerical simulation using continuum Landau-de Gennes modeling (see Methods). As shown in Fig.~2, the minimum energy configuration is a single disclination loop entangling the strip, in the location predicted by \eqref{eq:1}. Of course, the precise configuration depends on the strip's orientation relative to the far-field, but we have found that the orientation shown, with the strip's centerline in a plane perpendicular to the far-field orientation, has the lowest observed energy. The cross-section of the disclination loop shows a twisted $-1/2$ profile on the outside of the strip and a $+1/2$ twisted profile on the inside, as it has to in order to carry no hedgehog charge \cite{copar_chapter}. This hedgehog charge may be computed by several methods. The recently developed Pontryagin-Thom construction \cite{bryan} and methods related to disclination profile switching \cite{copar_chapter} both assert that the charge of the disclination is zero, as required on topological grounds. 

Perhaps the simplest generalization of the M\"{o}bius strip topology is to vary its embedding in $\mathbb{R}^3$. Different embeddings can be obtained by changing the number, $p$, of half-twists that the strip contains. The `canonical' M\"{o}bius strip has one half-twist ($p=1$), but more generally if $p$ is odd then the surface is still non-orientable and has the same topology as the M\"{o}bius strip. When $p$ is even the surface is orientable and has the topology of an annulus. Nonetheless, the embeddings are distinct and carry their own topological embellishments. The boundary of a M\"{o}bius strip is a circle. For a single half-twist this is a simple unknot, but for $p$ half-twists it is a $(p,2)$ torus knot ($p$ odd) or link ($p$ even). To see this, note that the boundary of a $p$-twisted strip ($p$ odd) lives on a torus whose major radius is that of the strip and whose minor radius is half the strip width. The curve the boundary draws on this torus goes round the meridional cycle $p$ times -- once for each half-twist -- while traversing the longitudinal cycle twice, which is the definition of a $(p,2)$ torus knot. The story is the same for orientable strips with $p$ even, except that there are two components to the boundary and they form a link. 

Can this structure, coming from the nature of the embedding, be exploited to controllably produce knotted and linked disclination loops in liquid crystals? Disclination lines that follow the surface of such a multiply-twisted M\"{o}bius strip will have the same shape and properties as the colloid boundary, yielding precisely constructed knots and links. Here we show that such configurations can be stabilized in chiral nematics; examples for doubly, triply, quadruply and quintuply twisted strips are shown in Fig.~3. They produce the $p=2$ Hopf link, the $p=3$ trefoil knot, the $p=4$ Solomon's knot and the $p=5$ cinquefoil knot, respectively. All these knots and links obey the topological requirements set out in the first section; the strips with an even number of twists are topologically trivial and the strip is entangled by an \textit{even} number of disclinations, those with an odd number of twists are non-orientable and so enclose an \textit{odd} number of disclinations.

The stablization of these knotted structures is not just a question of topology, energetics also come into play. In this regard, the chirality (inverse pitch) of the system has an important role in the stability of these configurations. Indeed it is generally true that chiral systems allow for more exotic structures~\cite{jampani11,toron}. In the achiral nematic system, the knotted defects are unstable and the liquid crystal assumes a ground state configuration consisting of $p$ small disclination loops entangling the strip along the contours where $\textbf{S}_n \cdot \textbf{n}_0=0$, as predicted by \eqref{eq:1}. A similar configuration, with slightly twisted loops (Fig.~4), is also the ground state in cholesterics -- the knots are metastable -- although the difference in energies is small (of order 1-2\%) and decreases both with increasing chirality and knot complexity $p$. The behavior with increasing $p$ can be understood in terms of the total length of disclination line, which scales as $p$ for the isolated loops and as $\sqrt{4+p^2}$ for the knots. If the chirality is increased such that the pitch becomes smaller than the width of the colloid then the disclinations develop twists analogous to those around spherical colloids~\cite{lintuvuori10,jampani11}. 

\begin{figure}[t]
\centering
\includegraphics[width=.49\textwidth]{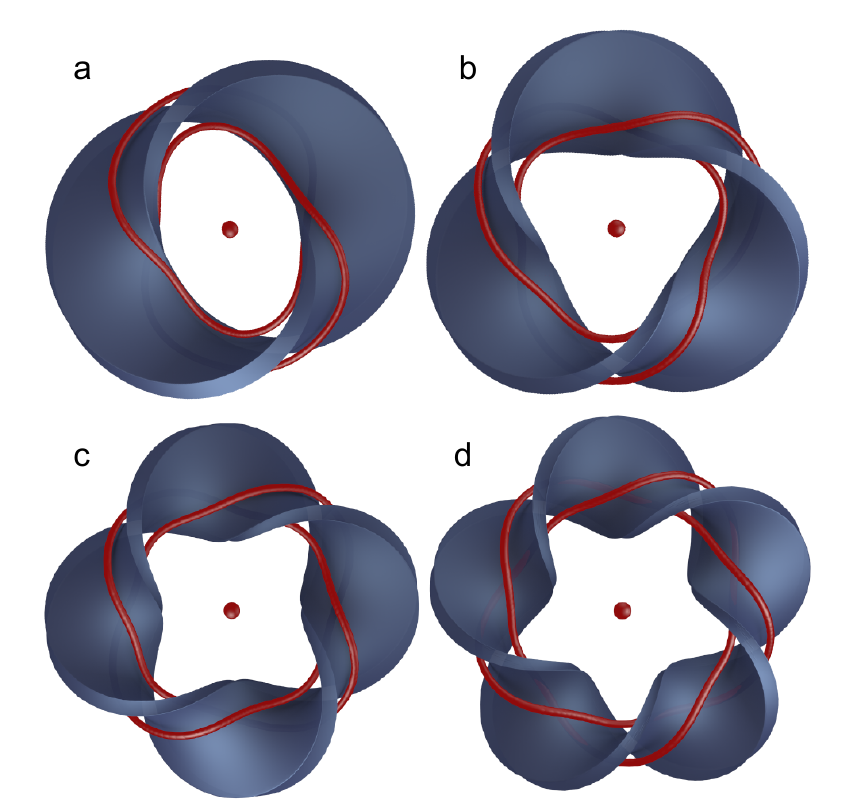}
\caption{Knotted and linked disclinations in chiral nematics stabilized by the presence of twisted surfaces with homeotropic boundary conditions on the flat faces, and planar on the thin edges. They are torus knots and links of the form $(p,2)$. Shown are: (a) $p=2$ Hopf link, (b) $p=3$ trefoil knot, (c) $p=4$ Solomon's knot and (d) $p=5$ cinquefoil knot. The defects in the center are hedgehogs, existing in pairs above and below the strip.}
\label{fig:knots}
\end{figure}

Like cholesterics, torus knots have a handedness -- the $(p,2)$ and $(p,-2)$ knots are mirror pairs -- and it is not surprising to find that the relative handedness influences the stability of the textures. The handedness of the knot is set by the shape of the colloid, and if this matches that of the cholesteric then knotted textures are stable, otherwise the system tries to expel the reverse twist in the configuration, resulting in an unstable knot.

The disclination lines themselves represent just a small portion of the entire system. Their knottedness imprints a complex orientational order on everything that is {\em not} the knot. Fig.~5a shows a slice through the director field of a quintuply twisted M\"{o}bius strip in a cell with fixed normal boundary conditions. To match the boundary conditions of the cell, the knotted disclination is accompanied by two hyperbolic hedgehogs, expanded into small loops, above and below the strip. The disclination-colloid pair has a constant profile that simply rotates uniformly as one moves around the strip; qualititavely, this structure does not depend on $p$, the number of half-twists. The cross-section through the colloid has a profile reminiscent of a double twist cylinder \cite{wright89}, split apart by the colloid into two separate $+1/2$ twist disclinations.

These configurations bear a striking similarity to the recently discovered `toron' textures \cite{toron}. Torons come in several flavors, but the ones that concern us have two hyperbolic defects above and below a double twist torus; that is, a double twist cylinder -- the building block of blue phases \cite{wright89} -- bent around such that it forms a torus. Like the knots of Fig.~3, these textures are stabilized in thin cells of chiral nematics, with uniform alignment at large distances. More than a superficial similarity, the knotted field configurations we present here can be thought of as the result of expanding the central part of a toron locally into two $+1/2$ twisted defects, that then rotate to form knots. The whole configuration is then stabilized by the colloid; indeed if the colloid is artificially removed from the simulation after equilibration, the configuration collapses into a toron. This connection may provide a potential route for the production of these textures using Laguerre-Gaussian polarizing beams.

\begin{figure}[t]
\centering
\includegraphics[width=.49\textwidth]{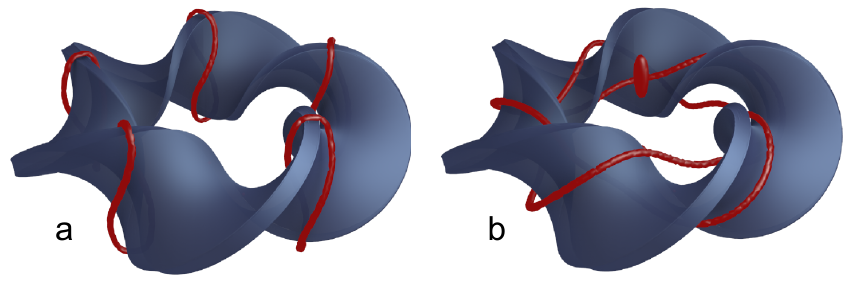}
\caption{Comparison of configurations for a quintuply twisted strip. (a) Groundstate configuration consisting of 5 small disclinations loops at the locations predicted by \eqref{eq:1}. They have a slightly twisted profile, caused by both the twisting of the strip and the chiral nature of the system. (b) $(5,2)$ torus knot entangling the strip, with two hyperbolic hedgehogs nucleated above and below. The energy difference between these configurations is approximately 1.2\%.}
\label{fig:rings}
\end{figure}

\begin{figure*}[t]
\centering
\includegraphics[width=.98\textwidth]{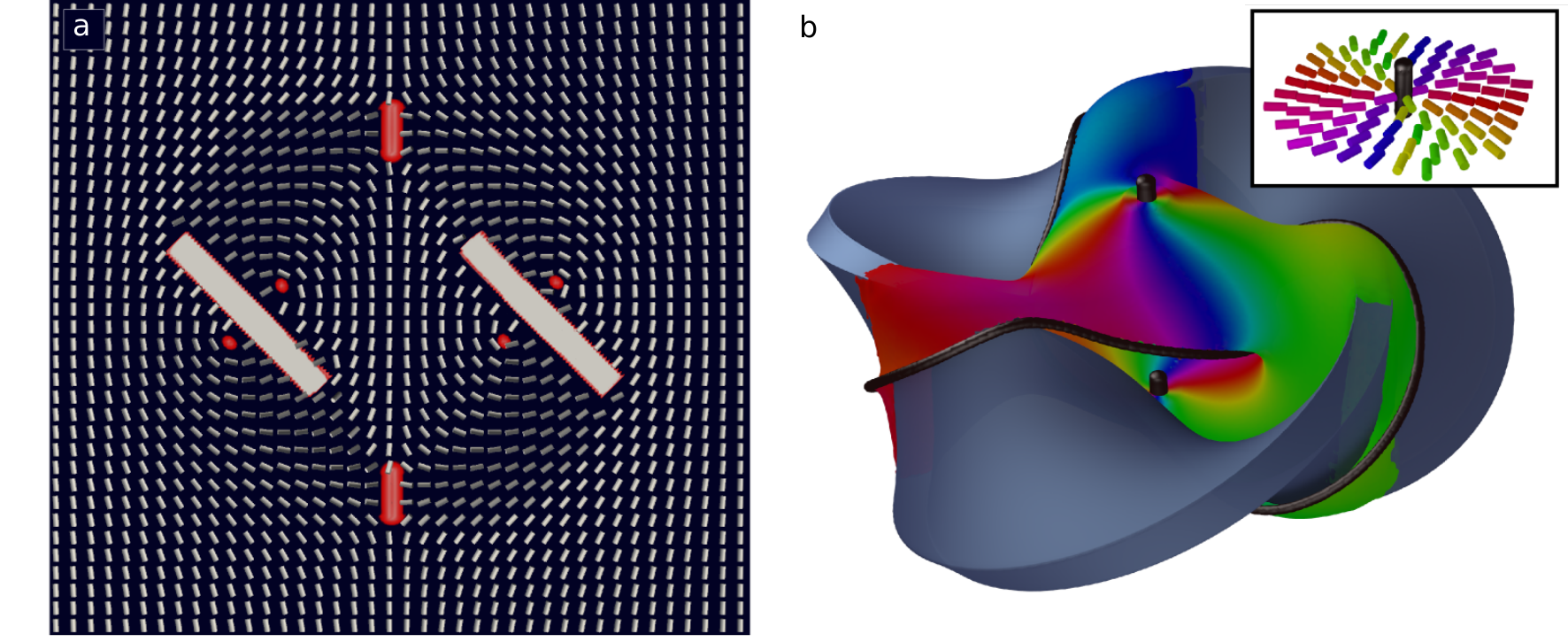}
\caption{Field Structure. (a): Director field profile of a $(5,2)$ torus knot and the two accompanying hyperbolic hedgehogs, opened up into small rings, above and below the strip, with disclinations bound to the colloid's surface. The local disclination-colloid profile is largely constant up to a rotation as one moves around the strip. Though this profile is for a $p=5$ configuration, qualitatively the director field does not change as one varies $p$. The far-field director is fixed to be vertical on the boundary of the cell. (b): Pontryagin-Thom surface~\cite{bryan} for the trefoil knot-colloid pair. The surface is constructed by considering all points where the director field is perpendicular to the far-field. The surface is then colored with the remaining $\mathbb{RP}^1$ degree of freedom (as illustrated in the inset). The defects, disclination line and two accompanying hedgehogs, are shown in black. The color winding around the hedgehogs establishes them as unit charge defects. The surface can be patched by removing the hedgehogs, and allowing the surface to pass through the colloid. One then obtains a Seifert surface for the knot, composed of two disks (top and bottom) connected by 3 (generally $p$) twisted bands. It is readily verified that this surface has genus 1 (generally $(p-1)/2$, for $p$ odd), the genus of the trefoil knot. While there is also a color winding around the disclination, this is a four-fold winding, giving the disclination even (equivalent to zero) hedgehog charge, as required.}
\label{fig:field}
\end{figure*}

A more quantitative investigation of the texture can be made through use of the Pontryagin-Thom (PT) construction \cite{bryan}, allowing us to examine the full topology of the field structure and even perform rudimentary `experimental knot theory'; we can use the texture around the knot to compute the knot genus, a classical knot invariant \cite{Lickorish}. The PT construction involves drawing the surfaces on which the director field is perpendicular to a given direction: such a surface is shown in Fig.~5b. Surfaces constructed this way must end either on topological defects or, as in our case, colloids. For a knotted disclination, this PT construction then yields a Seifert surface for the knot, and it is a classic result that the minimal genus over all possible Seifert surfaces is a knot invariant for the given knot \cite{Rolfsen}. To obtain a proper Seifert surface, the PT surface must be continued through the colloid and the points corresponding to the hedgehogs filled in\footnote{Since the hedgehogs are actually small loops in the simulations, an equivalent procedure is to take those loops as additional boundaries of a Seifert surface for the link consisting of the torus knot and two unlinked unknots. The genus of this Seifert surface is the same as that of the other one.}. Since our disclinations are torus knots, they have genus $(p-1)/2$ for $p$ odd and $(p-2)/2$ for $p$ even \cite{Rolfsen}. While, in principle, the genus of the PT surfaces is only bounded from below by this number, in the simulations presented here the limit is always reached, and we may read off the correct genus. 

In conclusion we have demonstrated the design and construction of elegant knotted and linked disclinations in liquid crystals, exploiting natural embeddings of twisted strips and a complete classification of the topological implications of homeotropic colloids. These knotted disclinations impose a complex structure on the director field, allowing knot invariants to be computed from the texture and fascilitating the full introduction of knot theory into experimental soft matter systems. The configurations themselves are intimately related to the recently discovered toron textures, suggesting a route for experimental realization. Future work will further explore the value of knot theory in understanding the subtleties of knotted fields and their potential for novel devices. 

%% == end of paper:

%% Optional Materials and Methods Section
%% The Materials and Methods section header will be added automatically.

%% Enter any subheads and the Materials and Methods text below.
\begin{materials}
To simulate the twisted strip colloids we use the standard Landau-de Gennes $Q$-tensor formalism. The liquid crystal order parameter is taken to be a traceless, symmetric tensor field $Q(\textbf{r})$. The free energy of the system, $F$, is then given as
\begin{equation}
\begin{split}
F & = \int_\Omega \biggl\{ \frac{A_0(1-\gamma/3)}{2}\, \text{tr}\bigl( {\bf Q}^2 \bigr) - \frac{A_0\gamma}{3}\, \text{tr}\bigl( {\bf Q}^3 \bigr) + \frac{A_0\gamma}{4} \Bigl( \text{tr}\bigl( {\bf Q}^2 \bigr) \Bigr)^2 \\
& \quad \qquad + \frac{L}{2}\, \text{tr} \Bigl( \bigl( \nabla\times{\bf Q} + 2q_0{\bf Q} \bigr)^2 \Bigr) \biggr\} \textrm{d}^3r \\ 
& \quad +\int_S \frac{W}{2}\, \text{tr}\Bigl( \bigl( {\bf Q}-{\bf Q}^0 \bigr)^2 \Bigr) \textrm{d}^2r.
\end{split}
\end{equation}
The domain $\Omega$ is the volume of the liquid crystal, and $S$ is the surface of the colloid. The constant $A_0$ defines the bulk energy scale of the material and $\gamma$ represents an effective temperature. $L$, the elastic constant, controls elastic distortions in the director field. $q_0$ is the chirality of the liquid crystal and is equal to $2\pi/P$ where $P$ is the cholesteric pitch. $q_0>0$ denotes a right-handed material and the pitch was set to be of order the colloid diameter, though the configurations were found to be stable at higher chiralities. Finally, $W$ controls the strength of the anchoring on the colloid surface, with the preferred orientation given by $Q^0$. The free energy is minimized by evolving the $Q$-tensor according to Landau-Ginzburg relaxational dynamics. These equations are then solved using finite difference methods on a cubic mesh of size $250\!\times\!250\!\times\!250$ or $260\!\times\!260\!\times\!150$. Simulation parameters ($A_0=0.16,\,L=0.03,\,\gamma=3.2,\,W=0.2$) were chosen to match typical values for a cholesteric liquid crystal with pitch comparable to the colloid radius (varied between 1 and 2 times this) and strong surface anchoring. 
\end{materials}

%% Optional Appendix or Appendices
%% \appendix Appendix text...
%% or, for appendix with title, use square brackets:
%% \appendix[Appendix Title]

\begin{acknowledgments}
We are grateful to Miha Ravnik for stimulating and beneficial discussions. This work was supported by the UK EPSRC. T Machon also supported by a University of Warwick Chancellor's International Scholarship. For their hospitality we thank the Physics department of the University of Ljubljana, where this work was completed.
\end{acknowledgments}

\end{article}

\end{document}